\documentclass[11pt]{article}

\usepackage{amsmath,amsfonts,amsbsy,amssymb,array,accents,dsfont}
\usepackage{enumerate,array,latexsym,graphicx,mathrsfs,verbatim,psfrag}
\usepackage{bm} 
\usepackage[normalem]{ulem}
\usepackage{booktabs} 
\usepackage[usenames]{color}
\usepackage[utf8]{inputenc}
\usepackage[english]{babel}
\usepackage{fancybox}

\usepackage{datetime}

\usepackage[nosort]{cite}
\usepackage{chngpage} 

\usepackage{enumitem}
\setlist{itemsep=0pt}

\usepackage[colorlinks=true,      linkcolor=darkblue,      urlcolor=darkblue,      
            filecolor=darkblue,      citecolor=darkblue,       pdfstartview=FitH,     
						pdfpagemode=UseNone,      bookmarksopen=true]{hyperref}  
\usepackage[all]{hypcap}     

\newcommand{\captionfonts}{\small}
\makeatletter  
\long\def\@makecaption#1#2{%
  \vskip\abovecaptionskip
  \sbox\@tempboxa{{\captionfonts #1: #2}}%
 \ifdim \wd\@tempboxa >\hsize
    {\captionfonts #1: #2\par}
  \else
    \hbox to\hsize{\hfil\box\@tempboxa\hfil}%
  \fi
  \vskip\belowcaptionskip}
\makeatother   

\DeclareMathSymbol{\medhatsym}{\mathord}{largesymbols}{"62} 

\DeclareMathSymbol{\medtildesym}{\mathord}{largesymbols}{"65}

\newcommand{\comm}[1]{} 

\def\IR{\mathbb{R}}

\def\({\left(}
\def\){\right)}
\def\[{\left[}
\def\]{\right]}

\def\coeff#1#2{{\textstyle \frac{#1}{#2}}}

\def\One{{\hbox{ 1\kern-.8mm l}}}

\def\barray{\begin{array}}
\def\earray{\end{array}}
\def\be{\begin{equation}}
\def\ee{\end{equation}}
\def\bea{\begin{eqnarray}}
\def\eea{\end{eqnarray}}
\def\bal{\begin{align}}
\def\eal{\end{align}}

\numberwithin{equation}{section} 

\makeatletter
\g@addto@macro\bfseries{\boldmath}
\makeatother

\definecolor{cardinal}{rgb}{0.6,0,0}
\definecolor{darkgreen}{rgb}{0,0.4,0}
\definecolor{purple}{rgb}{0.5, 0, 0.5}
\definecolor{golden}{rgb}{0.92, 0.7, 0}
\definecolor{midnight}{rgb}{0, 0, 0.5}
\definecolor{darkblue}{rgb}{0, 0, 0.8}

\def\coeff#1#2{\relax{\textstyle {#1 \over #2}}\displaystyle}

\def\IR{\mathds{R}}

\def\cA{{\cal A}}
\def\cB{{\cal B}}

\def\cF{{\cal F}}

\def\cN{{\cal N}}
\def\cO{{\cal O}}
\def\cP{{\cal P}}

\def\nBPS#1{$\frac{1}{#1}$-BPS}

\def\cO{{\cal O}}

\begin{document}


\begin{flushright}
%
%
\end{flushright}

\vspace{14mm}

\begin{center}

{\huge \bf{Early Scrambling and Capped BTZ Geometries}} \medskip \\


\vspace{13mm}

\centerline{{\bf  Iosif Bena$^1$, Emil J. Martinec$^2$, Robert Walker$^{3}$ and Nicholas P. Warner$^{3,4}$}}
\bigskip
\bigskip
\vspace{1mm}

\centerline{$^1$ Institut de Physique Th\'eorique,}
\centerline{Universit\'e Paris Saclay, CEA, CNRS, }
\centerline{Orme des Merisiers,  F-91191 Gif sur Yvette, France}
\bigskip
\centerline{$^2$Enrico Fermi Inst.\ and Dept.\ of Physics, }
\centerline{University of Chicago,  5640 S. Ellis}
\bigskip
\centerline{$^3$\,Department of Physics and Astronomy,}
\centerline{University of Southern California,} \centerline{Los
Angeles, CA 90089-0484, USA}
\bigskip
\centerline{$^4$\,Department of Mathematics,}
\centerline{University of Southern California,} \centerline{Los
Angeles, CA 90089, USA}

\vspace{4mm}

%

\vspace{8mm}
 
\textsc{Abstract}

\begin{adjustwidth}{10mm}{10mm} 
 %
\vspace{3mm}
\noindent

Geodesic probes in certain horizonless microstate geometries experience extreme tidal forces long before reaching the region where these geometries differ significantly from the extremal BTZ black hole. The purpose of this paper is to show that this behavior is a universal feature of all geometries that have a long BTZ throat that terminates in a cap, regardless of the details of this cap. Hence, incoming probes will scramble into the microstate structure before they encounter the region where the charges of the solution are sourced, and the reason for this premature scrambling is the amplification of tiny geometrical deviations by the relativistic speeds of the probes.    
To illustrate the  phenomenon, we construct a new family of smooth horizonless superstratum microstate geometries, dual to D1-D5 CFT states whose momentum charge is carried by excitations on CFT strands of length $k$.   We also show that, in the large-$k$ limit, these new superstrata resemble a blackened supertube solution everywhere except in the near-supertube region. Thus they resolve the singularity caused by the naive back-reaction of modes with non-linear instabilities near evanescent ergosurfaces.

%
\end{adjustwidth}

\end{center}


\thispagestyle{empty}

\newpage


\baselineskip=14pt
\parskip=2pt

\tableofcontents


\baselineskip=15pt
\parskip=3pt

\section{Introduction}
\label{Sect:introduction}

\subsection{Microstate geometries}
\label{Sect:MGs}

Microstate geometries are defined to be smooth, horizonless geometries that have the same asymptotic charges as a particular black hole or black ring.  The first examples of microtstate geometries were ``shallow'' in that they corresponded to BPS black holes and black rings with vanishingly small horizon areas and so did not have deep AdS$_2$ throats \cite{Bena:2005va,Berglund:2005vb, Bena:2006is}.  Shortly thereafter, the first ``deep'' microstate geometries were constructed, corresponding to black holes and black rings with macroscopic horizon areas \cite{Bena:2006kb,Bena:2007qc,Bena:2007kg}. These geometries, which correspond, upon dimensional reduction, to certain classes of four-dimensional scaling multi-center solutions \cite{Denef:2007vg} have an arbitrarily long AdS$_2$ throat whose depth, before capping off, is determined by a (classically) free parameter.  However, these examples of geometries typically had a parametrically large angular momenta in five dimensions, and hence corresponded only to BMPV black holes close to the cosmic censorship bound.

The final step in finding microstate geometries at generic points in the phase space of black holes was achieved in the last two years with the construction of microstate geometries that have arbitrarily small angular momenta \cite{Bena:2016ypk,Heidmann:2017cxt, Bena:2017fvm, Avila:2017pwi, Bena:2017xbt}. These microstate geometries can be written in many duality frames, but the most convenient is the IIB on T$^4$ (or K3) in which the charges of the black hole correspond to D1 and D5 branes with momentum along their common direction. In this frame it is possible to take an AdS$_3$ near-horizon limit of these microstate geometries (essentially by removing the constants in the harmonic functions corresponding to the D1 and D5 charges), and the resulting solutions have three scales: an asymptotic region, in which the D1 and D5 charges dominate the geometry and the metric is approximately AdS$_3$ $\times S^3$. Closer in, the momentum charge begins to make its presence felt and stabilizes the size of the $AdS_3$ circle, so that the metric becomes that of AdS$_2$ $\times S^1 \times S^3$.  In the generic classes of solutions the depth of the AdS$_2$ throat is governed by the $J_R$ angular momentum of the solution\footnote{In our convention, supersymmetric black holes can have a non-zero $J_L$ but must have zero $J_R$} and becomes infinitely deep as $J_R \to 0$.  For small, but  non-zero $J_R$, the radius of the  $S^1$ eventually starts to shrink again, and the throat caps off smoothly. 

The AdS$_3$  $\times S^3$ and the AdS$_2$ $\times S^1 \times S^3$ regions (which we will refer from now on as the  AdS$_3$ and the AdS$_2$ regions) are extremely closely approximated by the six-dimensional uplift of an extremal BTZ black hole. Hence, the deep scaling microstate geometries may be thought of as {\it smoothly capped, BTZ geometries}.

\subsection{Capping BTZ geometries}
\label{Sect:IntroCappedBTZ}

We can also define a more general {\it  capped, BTZ geometry} to be any background that has the same AdS$_3$ and AdS$_2$ regions as an extremal BTZ black hole, but with some non-trivial structure ``at the bottom'' of the AdS$_2$  throat.  This structure necessarily breaks conformal invariance. The microstate geometries are, of course, {\it smoothly capped BTZ geometries}, and they are the most relevant for understanding the physics of the corresponding black hole. However, if we do not require smoothness we can create capped BTZ geometries in many ways: we can separate  the D1 branes and D5 branes and we can distribute the momentum charge as we wish.  We can also combine the D1 and D5 branes into a two-charge supertube \cite{Lunin:2001fv, Mateos:2001qs, Emparan:2001ux} and then add a source for the  momentum either exactly on the supertube, or in some distribution around it. We can also consider geometries with multiple black holes and black rings located at the bottom of a long AdS$_2$ throat. 

The first purpose of this paper is to show that the broad class of capped BTZ  geometries have the same universal effects on infalling geodesic probes as the much more complicated, smoothly-capped microstate geometries. Hence, for the purpose of understanding when, and how, infalling probes are scrambled by the black hole microstates, one does not need to use all the details of the smooth-capped BTZ geometries, but just some of its generic features that are also captured by the more-easily constructed, singular, capped BTZ geometries. This simplification is very useful, especially if one wants to analyze whether microstate geometries can give different signatures in gravity waves than the corresponding black hole. Analyzing each and every such geometry would be arduous, and therefore, identifying universal features that distinguish microstate geometries from the extremal BTZ solution and are captured by the relatively-simple singular, capped BTZ geometries is a very valuable simplification.

The second purpose of the paper, which is technically independent but touches upon the same physics, is to construct a new class of superstratum microstate geometries. These microstate geometries, whose dual states at the orbifold point of the D1-D5 CFT are well understood, can be viewed as more examples of smooth capped BTZ geometries. However, what distinguishes them from the previously-constructed microstate geometries is that they have a limit in which one can see how their geometries approach a simple, but singular,  capped BTZ geometry. Hence, these microstate geometries can be thought of as resolutions of the singularity caused by the naive back-reaction of modes with non-linear instabilities near evanescent ergosurfaces discussed in \cite{Eperon:2016cdd,Marolf:2016nwu}.

\subsection{Overview of our results}
\label{Sect:overview}

The tidal forces on geodesic probes of microstate geometries were first investigated in \cite{Tyukov:2017uig}, where it was shown that such tidal forces become large ({\it i.e.}~approach the Planck scale) long before the probe encounters the cap at the bottom of the throat. Indeed, the tidal forces typically become large ``half-way down'' the throat:  at the geometric mean of the scales at the top and bottom of the $AdS_2$ BTZ-like throat.   This was a surprise because, like the extremal BTZ geometry, the Riemann invariant is everywhere limited by the D1 and D5 charges, $N_1$ and $N_5$, and vanishes as $(N_1 N_5)^{-1}$ as these charges become large.  The large tidal forces arise because capped BTZ geometries have very small deviations from the BTZ geometry in the throat but these are greatly magnified by the relativistic motion of the probe.   We will argue that this is a universal property of generic capped BTZ geometries. Indeed, while the expressions for tidal forces on probes can be very complicated and receive many contributions, we find that for probes falling from the top of the throat in a generic capped BTZ geometry, there is a tidal-force term whose magnitude is:
\begin{equation}
|\cA|_{\rm throat}  ~\sim~ \frac{a^2  \, Q_P^2 }{\sqrt{Q_1 Q_5} \,  r^{6}}   \,.
  \label{cAgeneric}
\end{equation}
In this equation, $Q_1$ and  $Q_5$ are the underlying supergravity charges of the D1 and D5 branes, $Q_P$ is the momentum charge,  $a$ is the scale of the cap  (determined by the residual  angular momentum of the solution)  and $r$ is the location of the probe particle with $a \ll r \ll \sqrt{Q_P}$.  
This is the term that generates the large tidal forces in the throat.  It is important that this is not only proportional to $Q_P^2$ but is also proportional to $a^2$.  This means that this term vanishes in the BTZ limit, $a \to 0$.  Thus the large tidal forces are a result of the depth of the throat, controlled by $Q_P$ {\it and the presence of the cap.}

Our original motivation for constructing a new family of superstratum microstate geometries was to explore geometries whose CFT duals suggested that the probes might experience a somewhat softer infall.   In particular, our new family of microstate geometries have an integral parameter, $k$, that represents the length of the CFT strands that carry the momentum charge.  Since the CFT momentum-gap  of these momentum-carrying sectors scales as $1/k$, a microstate geometry that is dominated by a large-$k$ sector might afford a ``softer impact'' for incoming probes.%
\footnote{It is important to remember that such an intuition is based on the symmetric orbifold CFT and this does not always match that of the physics of the geometry, since the orbifold locus in the moduli space lies far from the regime where supergravity is the appropriate effective description. Moreover, tidal forces on non-BPS probes are very unlikely to be ``protected''  under deformations of the moduli.} On the contrary, the scattering of probes in these new families led to the discovery of the universal  expression (\ref{cAgeneric}) for capped BTZ geometries: a result that does not involve any details of the CFT state.  

We will see that, for large $k$, our new microstate geometries limit to blackened supertube geometries, and that there are vast tidal forces in the near-supertube region.  This leads to a new perspective on the possible end-state for the non-linear instabilities discussed in \cite{Eperon:2016cdd,Marolf:2016nwu}.  In these papers it was argued that infalling matter would be trapped for long periods of time in the neighborhood of evanescent ergosurfaces.  Such surfaces are defined  as the loci upon which the Killing vector that is time-like at infinity, becomes null (these surfaces are evanescent in that the Killing vector  never becomes space-like).  For the superstratum microstate geometries constructed by adding momentum to a supertube \cite{Bena:2011uw, Bena:2015bea,Bena:2016ypk,Bena:2017xbt}, the evanescent ergosurface is precisely the supertube locus. Thus infalling matter is expected to be trapped for long periods of time in the neighborhood of the original supertube.   

 Trapping of matter around evanescent ergosurfaces was already anticipated a decade ago in the microstate geometry program, albeit in another guise known as {\it entropy enhancement}.  It was observed in \cite{Bena:2008nh} that, for a given set of asymptotic charges, fluctuating supertubes localized on, or near, an evanescent ergosurface can give rise to vastly more entropy than they can give when placed at a generic region in the space-time.  Put simply, there is a lot more free energy available in the neighborhood of an evanescent ergosurface. This is in close accord with the observation in  \cite{Eperon:2016cdd} that states in the neighborhood of evanescent ergosurfaces have vanishingly small energies when measured from infinity.  It therefore follows that near-extremal black-hole microstate structure will accumulate around evanescent ergosurfaces\footnote{Such a classical accumulation of matter is, of course, limited by quantum mechanics for much the same reason that electrons do not accumulate in atomic nuclei. However, the quantization of merely the supertube degrees of freedom  \cite{Bena:2008nh,Palmer:2004gu}  suggests that there is still a large amount of free energy associated with evanescent ergosurfaces.}.

Trapping matter for long periods of time is also something one should expect of a microstate geometry if it is to replicate the classical behavior of a black hole.  The potential problem is that accumulating matter in a small region could lead to the formation of singularities in the form of black rings, black holes or shock waves  \cite{Eperon:2016cdd,Marolf:2016nwu}.  While the formation of black objects is a natural outcome in general relativity, the whole point of the fuzzball and microstate geometry programmes is that string theory comes with vastly more degrees of freedom that enable highly collapsed states of matter to avoid the inevitable formation of singularities and horizons.  
Indeed, these programmes are trying to create a paradigm shift in which one takes the view that singular geometries and horizons are artifacts of restricting the degrees of freedom to simple gravitational configurations in low dimensions, and that black-hole microstates are best described by the vast range of microstate geometries, microstate solutions and fuzzballs afforded by string theory \cite{Bena:2013dka}\footnote{More broadly, it also remains a distinct possibility that the notion of horizon and/or singularity depends on the probe under consideration -- for instance that supergravitons see horizons and singularities while stringier entropy-encoding degrees of freedom see neither.}.

To this end, rather than reaching for the old tool-kit of black objects and shock waves, one should instead look for fuzzballs and microstate geometries that would better represent the end states of trapped matter near evanescent ergosurfaces.   Indeed, it was argued in \cite{Marolf:2016nwu} that the trapped matter is likely to cascade into higher and higher frequency modes ultimately requiring string theory, rather than supergravity, for its proper description.

While generic microstate solutions are expected to be intrinsically stringy, suitably coherent forms of such microstate structure should appear as microstate geometries.  We therefore find it very gratifying that the new class of solutions obtained in this paper can be viewed as resolving a singular blackened supertube solution into a microstate geometry.  While the geometries described here are dual to very special coherent states, and we are by no means claiming that they are generic elements of the black hole ensemble, our result shows that a singular configuration created by the naive back-reaction  at the evanescent ergosurface of a supertube can be resolved into a simple, smooth horizonless geometry that looks exactly like the singular object until one is very close to the supertube locus. While we have not proven that this will always happen, the new example presented in this paper  demonstrates, in principle, that the microstate geometry/fuzzball paradigm is alive and well in the vicinity of evanescent ergosurfaces.

More generally, one  may hope that a sufficiently detailed analysis in string theory will reveal the entropic constituents of black holes as seen from the gravitational side of gauge/gravity duality.  Supertubes near the evanescent ergosurface get us partway there  \cite{Bena:2008nh,Warner:2008ma}, and recent investigations of supertubes in string theory \cite{Martinec:2017ztd,Martinec:2018nco} have shown how the long string structure of the non-gravitational dual begins to emerge near the extremal black hole transition.  One might hope that one could merge these various threads of investigation to provide a more complete picture of how fuzzballs form and develop. 

In Section  \ref{sec:CappedBTZ} of this paper we will discuss more precisely what we mean by capped BTZ geometries and describe several examples.  In Section \ref{sec:MGs} we describe the new family  of  microstate geometries, indexed by the length, $k$, of the strands that are carrying the momentum charge.  We also show that, in the large-$k$ limit, these microstate geometries look very much like blackened supertubes until one is extremely close to the supertube locus.    In Section \ref{sec:Geodesics} we study a class of geodesic probes in our microstate geometries and in capped BTZ geometries more generally.  Our analysis  closely parallels that of  \cite{Tyukov:2017uig} but we extract the universal term (\ref{cAgeneric}) responsible for the large tidal forces in any capped BTZ geometry.  Finally,  Section  \ref{sec:Disc} contains some brief concluding remarks.

\section{Capped BTZ geometries} 
\label{sec:CappedBTZ}

We consider supersymmetric solutions of Type IIB string theory on T$^4$ or K3, whose six-dimensional part can be written as a solution of six-dimensional supergravity with a metric\cite{Gutowski:2003rg}:
\begin{equation}
d s^2_{6} ~=~-\frac{2}{\sqrt{\cP}}\,(d v+\beta)\,\Big[\, d u+\omega + \coeff{1}{2} \,\mathcal{F}\, (d v+\beta)\, \Big ]+\sqrt{\cP}\,d s^2_4\,.
\label{sixmet}
\end{equation}
For the simplest families of solutions,  the metric, $ds_4^2$, on the four-dimensional base, $\cB$, is simply that of flat $\IR^4$, which we write  in terms of spherical bipolar coordinates:
 \begin{equation}
 d s_4^2 ~=~ \Sigma \, \left(\frac{d r^2}{r^2+a^2}+ d\theta^2\right)+(r^2+a^2)\sin^2\theta\,d\varphi_1^2+r^2 \cos^2\theta\,d\varphi_2^2\,,
 \label{ds4flat}
\end{equation}
where
 \begin{equation}
\Sigma~\equiv~  (r^2+a^2 \cos^2\theta)     \,.
 \label{Sigdefn}
\end{equation}
The coordinates, $u$ and $v$, are the standard null coordinates, which are related to the canonical time and spatial coordinates via:
\begin{equation}
  u ~=~  \coeff{1}{\sqrt{2}} (t-y)\,, \qquad v ~=~  \coeff{1}{\sqrt{2}}(t+y) \,, \label{tyuv}
\end{equation}
where $y$ is the coordinate around $S^1$ with
\begin{equation}
  y ~\equiv~  y ~+~ 2\pi  R_y \,. \label{yperiod}
\end{equation}

We will work within the six-dimensional supergravity coupled to two tensor multiplets and, since the graviton supermultiplet contains a tensor gauge field, we have three tensor gauge fields.  These fields, and the scalars in the supermultiplets, are determined by scalar potentials, $Z_I$, and magnetic two-form fields, $\Theta_I$, on the four dimensional base, $\cB$ \cite{Bena:2011dd,Bena:2015bea}.  For historical reasons\footnote{These solutions was originally formulated in five dimensions and the fields $Z_3$ and  $\Theta_3$ have become part of the Kaluza-Klein geometry: $Z_3$ has essentially become $\cF$ in (\ref{sixmet}) and $\beta$ is the potential for $\Theta_3$.} the index $I$ takes the values $1,2,4$.     

The potentials, $Z_1$, $Z_2$ and $\cF$, encode the electric D1, D5 and momentum (P) charges while the  $\Theta_I$  encode magnetic dipoles.  In the following $Z_4$ and  $\Theta_4$ will either vanish or will only have dipole, or higher multipole, moments.

Supersymmetry requires the warp factor, $\cP$, to be determined in terms of the structure constants and scalar potentials of the supergravity: 
\begin{equation}
\cP ~\equiv~   Z_1\,Z_2 -  Z_4^2 \,. \label{Pform}
\end{equation}
%

\subsection{The D1-D5 supertube}
\label{ss:STs}

The simplest archetype of a capped geometry is that of a D1-D5 supertube.  This, arguably, was the start of the fuzzball program, in which it was realized that a two-charge black hole can be  resolved into a smooth geometry.  The supertube has a KKM dipole charge, and only D1 and D5 charges. The corresponding warp factors are:
\begin{equation}
Z_1 ~=~ \frac{Q_1}{\Sigma}  \,, \qquad Z_2 ~=~ \frac{Q_2}{\Sigma} \,, \qquad   Z_4 ~=~ 0  \,, \qquad   \cF ~=~ 0 \,,
\label{STelectric}
 \end{equation}
and the KKM dipole charge is comes from the non-trivial fibration vector, $\beta$: 
 \begin{equation}
\beta ~=~  \frac{R_y \, a^2}{\sqrt{2}\,\Sigma}\,(\, \sin^2\theta\, d\varphi_1 - \cos^2\theta\,d\varphi_2\,)   \,.
 \label{betadefn}
\end{equation}
The supertube also has a nontrivial angular momentum:
\begin{equation}
\omega ~=~ \omega_0 \,, \qquad \omega_0 ~\equiv~  \frac{a^2 \, R_y \, }{ \sqrt{2}\,\Sigma}\,  (\sin^2 \theta  d \varphi_1 + \cos^2 \theta \,  d \varphi_2 ) \,.
\label{angmom0}
\end{equation}

The supertube is located at  $\Sigma = 0$, or $r = 0$, $\theta =\frac{\pi}{2}$.   Regularity at the supertube  imposes the constraint:
\begin{equation} 
Q_1Q_2 ~=~R_y^2 \,  a^2  \,.
\label{STreg}
\end{equation} 
From (\ref{sixmet}), one sees that the $u$ and $v$ directions in (\ref{sixmet}) become null when $\cP \to \infty$, and so the evanescent ergosurfaces are located at the supertube locus $r = 0$, $\theta =\frac{\pi}{2}$.  

The electric charges $Q_1, Q_2 \equiv Q_5$ are related to the quantized charges, $N_1$ and $N_5$,  via \cite{Bena:2015bea}:
\begin{equation}
Q_1 ~=~  \frac{(2\pi)^4\,N_1\,g_s\,\alpha'^3}{V_4}\,,\quad Q_5 = N_5\,g_s\,\alpha'\,,
\label{Q1Q5_N1N5a}
\end{equation}
where $V_4$ is the volume of $T^4$ in the IIB compactification to six dimensions.  In particular, it is convenient to define $\cN$ via: 
\begin{equation}
\cN ~\equiv~ \frac{N_1 \, N_5\, R_y^2}{Q_1 \, Q_5} ~=~\frac{V_4\, R_y^2}{ (2\pi)^4 \,g_s^2 \,\alpha'^4}~=~\frac{V_4\, R_y^2}{(2\pi)^4 \, \ell_{10}^8} ~=~\frac{{\rm Vol} (T^4) \, R_y^2}{ \ell_{10}^8} \,,
\label{cNdefn}
\end{equation}
where $\ell_{10}$ is the ten-dimensional Planck length and  $(2 \pi)^7 g_s^2 \alpha'^4  = 16 \pi G_{10} ~\equiv~ (2 \pi)^7 \ell_{10}^8$.    The quantity, ${\rm Vol} (T^4)  \equiv (2\pi)^{-4} \, V_4$, is sometimes introduced \cite{Peet:2000hn} as a ``normalized volume'' that is equal to $1$ when the radii of the circles in the $T^4$ are equal to one in Planck units.

The left- and right-moving angular momenta of the supertube are given by: 
\begin{equation}
J_L ~=~  J_R~=~    \frac{1}{2} \, \cN  \,a^2 \, ,
\label{st-angmom}
\end{equation}
and their equality reflects the fact that the supertube only rotates inside one of the $\IR^2$s inside the $\IR^4$ base.

For this solution, one  can rewrite the metric, (\ref{sixmet}), as global AdS$_3\times S^3$:
\begin{equation}
\begin{aligned}
d s^2_{6} ~=~  \sqrt{Q_1 Q_5}\, \bigg[  \, &   -\frac{ (r^2+ a^2)}{a^2 \, R_y }\,dt^2 ~+~  \frac{dr^2}{(r^2+ a^2)} ~+~  \frac{ r^2}{a^2 \, R_y }\,dy^2\\
&+ ~ d\theta^2 ~+~ \sin^2 \theta \, \bigg(d\varphi_1 - \frac{1}{R_y }\,dt\bigg)^2~+~ \cos^2 \theta \,\bigg(d\varphi_2 - \frac{1}{R_y }\,dy\bigg)^2   \bigg] \,.
\end{aligned}
\label{STmet}
\end{equation}

Thus the cap is simply the bottom of the global AdS$_3$.  The geometry is asymptotic to AdS, and not flat space, because we did not add any constants to the electrostatic potentials in (\ref{STelectric}).  

\subsection{The extremal BTZ geometry}
\label{ss:BTZ}

The extremal  BTZ metric is obtained  by adding a momentum charge and taking $a \to 0$ in  (\ref{ds4flat}):
\begin{equation}
 Z_1~=~ \frac{Q_1}{r^2} \,, \quad  Z_2~=~ \frac{Q_2}{r^2} \,, \quad Z_4 ~=~ 0\,;  \qquad \cF ~=~ -  \frac{2\,Q_P}{r^2}\,, \qquad \beta ~=~ \omega ~=~ 0    \,.  
  \label{BTZsol}
\end{equation}
The factor of $-2$ in $\cF$ comes from the fact that, for asymptotically AdS superstrata, the most natural identification is to write  the electrostatic momentum potential as  $Z_3 = 1 -\frac{1}{2} \cF$. (See, for example, \cite{Bena:2015bea}.)

The six-dimensional metric may then be written as the extremal BTZ metric with a round $S^3$:
\begin{equation}
 ds_6^2  ~=~ \sqrt{Q_1 Q_2} \,  \bigg(  \bigg[ \, \frac{d\rho^2}{\rho^2}  - \rho^2 \, dt^2 +  \rho^2 \, dy^2 + \frac{Q_P}{Q_1 \, Q_2} \, (dy+dt)^2 \bigg] 
~+~
\big[  d\theta^2  +  \sin^2 \theta \, d\varphi_1^2 + \cos^2 \theta  \, d\varphi_2^2 \, \big] \bigg)  \label{BTZmet} \,,
\end{equation}
where $\rho \equiv (Q_1 Q_2)^{-\frac{1}{2}} \,r$.  

This solution has  charges $Q_1, Q_2$ and $Q_P$ but no angular momentum. 

For $r > \sqrt{Q_P}$, this geometry is AdS$_3$ $\times S^3$ and for $r <\sqrt{Q_P}$, the radius of the $y$-circle  becomes fixed and one enters the infinitely-long BTZ throat.  The limit $r \to 0$ corresponds to approaching the horizon of the black hole.

\subsection{A blackened supertube and capped BTZ geometries}
\label{ss:BSTs}

The simplest BPS solution that caps off and carries momentum charge is a blackened supertube.  That is, one  adds momentum charge along the supertube locus at $r = 0$, $\theta =\frac{\pi}{2}$:  
\begin{equation}
 Z_1~=~ \frac{Q_1}{\Sigma} \,, \quad  Z_2~=~ \frac{Q_2}{\Sigma} \,, \quad Z_4 ~=~ 0\,;  \qquad \cF ~=~ -  \frac{2\,Q_P}{\Sigma}   \,.  
   \label{BSTcharges}
\end{equation}

The fibration vector,  $\beta$, is given by (\ref{betadefn}) and so there is now a non-trivial source term, $\cF d \beta$, in the  BPS equations for $\omega$.  The exact BPS solution is given by:
\begin{equation}
 \omega ~=~  \omega_0  ~+~ \sqrt{2}\, a^2\, Q_P \,  R_y \, \frac{ \sin^2 \theta  \, \cos^2 \theta }{\Sigma^3}\, \big[ \, (r^2 +a^2)\, d \varphi_1~-~ r^2  \, d \varphi_2 \,\big] \,.
  \label{BSTomega}
\end{equation}
where $\omega_0$ is given by (\ref{angmom0}).  

This solution has  charges $Q_1, Q_2, Q_P$ and angular momenta $J_L = J_R = \frac{1}{2} \cN a^2$. However, one should note that as $\Sigma \to 0$, $\omega$ diverges as $\Sigma^{-3}$ and the leading terms  in the metric in the $(\varphi_1, \varphi_2)$ directions are: 
\begin{equation}
\begin{aligned}
-\frac{2}{\sqrt{\cP}}\, \beta \,  \omega ~=~ & -  \frac{2\, a^4\, Q_P \,  R_y^2}{Q_1 Q_2} \, \frac{ \sin^2 \theta  \, \cos^2 \theta }{\Sigma^3}\,  ( (r^2 +a^2)\, d \varphi_1~-~ r^2  \, d \varphi_2 ) ( \sin^2\theta\, d\varphi_1 - \cos^2\theta\,d\varphi_2)  \\
 ~\sim~ & - \frac{2\, a^6\, Q_P \,  R_y^2}{\sqrt{Q_1 Q_2}} \, \frac{ \sin^2 \theta  \, \cos^2 \theta }{\Sigma^3}\,  d \varphi_1^2  \,, 
\end{aligned}
  \label{leadangles}
\end{equation}
as $r \to 0$, $\theta \to \frac{\pi}{2}$.  There is, of course, an order of limits issue because the numerator has a factor of $ \cos^2 \theta$, but taking $r \to 0$ faster than $\theta \to \frac{\pi}{2}$ makes this term strongly divergent.

This means that this rather simple blackening of the supertube has created  closed time-like curves in the immediate neighborhood of the light-like loop of the original supertube.   In spite of having such a pathology, this metric will be very useful in our broader discussion because we want to consider metrics that look very much like this metric until one gets close to the supertube region, $r  \lesssim a$. 

We also note in passing that there are blackened supertubes that do not have closed time-like curves but these require adding at least one more  magnetic dipole charge \cite{Bena:2004wv}.  As we will discuss in Section \ref{sec:Disc}, such solutions will be relevant to generalizations of the work presented here. 

In the limit  $a \to 0$,  this metric becomes that of the extremal BTZ black hole (\ref{BTZmet}).  This geometry therefore generically has a long BTZ throat whose depth is  controlled by $a^{-1} \sqrt{Q_P}$.    This is an archetype of the capped BTZ geometry in that it has three regions: (i) AdS$_3$ $\times S^3$  for $r > \sqrt{Q_P}$; (ii) A long BTZ throat for $a < r<  \sqrt{Q_P}$,  and a cap that looks like global AdS$_3$ for  $r < a$ and $\theta < \frac{\pi}{2}$.   The metric described here does, of course, have a singularity and closed time-like curves at $r=0$, $\theta = \frac{\pi}{2}$ but so long as one stays away from the momentum-charge sources, the geometry caps off much like global AdS$_3$.  In Section  \ref{sec:MGs} we will discuss completely smooth microstate geometries that have this capped BTZ structure.

The importance of this capped BTZ geometry is that it is the simplest geometry that illustrates how the presence of the cap has a very significant effect on tidal forces quite high up the BTZ throat.  Indeed,  for $a \ll r \ll \sqrt{Q_P}$, some of the dominant tidal properties  can be captured by taking:  
\begin{equation}
 Z_1~\simeq~ \frac{Q_1}{r^2} \,, \quad  Z_2~\simeq~ \frac{Q_2}{r^2} \,, \quad Z_4 ~=~ 0\,;  \qquad \cF ~\simeq~ -  \frac{2\,Q_P}{r^2}   \,, 
   \label{UninversalCharges}
\end{equation}
but with
\begin{equation}
 \beta ~=~  \frac{R_y \, a^2}{\sqrt{2}\,\Sigma}\,(\, \sin^2\theta\, d\varphi_1 - \cos^2\theta\,d\varphi_2\,) \,, \qquad \omega ~=~  \omega_0  ~=~    \frac{a^2 \, R_y \, }{ \sqrt{2}\,\Sigma}\,  (\sin^2 \theta  d \varphi_1 + \cos^2 \theta \,  d \varphi_2 ) \,.
  \label{Uninversalvecs}
\end{equation}
This does not satisfy the BPS equations exactly, but this is an extremely good approximation to capped BTZ geometries in the throat where one has $a \ll r \ll \sqrt{Q_P}$.  In particular,  choosing  (\ref{UninversalCharges}) creates the BTZ throat in a global AdS geometry while using  (\ref{Uninversalvecs}) retains enough details of the cap to influence the geometry and capture the tidal forces that will be of interest in Section~\ref{sec:Geodesics}.  We will therefore think of (\ref{UninversalCharges}) and (\ref{Uninversalvecs}) as describing a universal capped BTZ geometry built by adding momentum charge in and around an underlying supertube structure.

\section{The microstate geometries} 
\label{sec:MGs}

\subsection{The CFT states and dual geometries}
\label{ss:CFTstates}

Following \cite{Tyukov:2017uig},  we focus on geometries that are holographically dual to the pure momentum excitations of the D1-D5 system.  The numbers of underlying D1-branes and D5-branes will be denoted by $N_1$ and $N_5$ respectively.  The ground states of the D1-D5 system can be described by partitioning $N= N_1 N_5$ into strands of lengths between $1$ and $N$ and by the polarizations of each of the strands (see, for example,\cite{Bena:2015bea}).  One can then excite the strands using operators in the CFT and the standard set of \nBPS{8} states are those that remain in the right-moving Ramond ground state but have arbitrary excitations in the left-moving sector.    Here we construct the geometries that are dual to coherent superpositions of states assembled from strands of length $1$ with polarization $|\!+\!+\rangle$, and a set of strands of length $k$ with polarization $|00\rangle$. For simplicity we also only consider a single, left-moving excitation obtained by acting once with $(L_{-1}- J^3_{-1})$ on the $|00\rangle$ strands.  That is we consider the states:
 \begin{equation}
(|\!+\!+\rangle_1)^{N_{++}} \big( (L_{-1}- J^3_{-1}) \, |00\rangle_k \big)^{N_{00}}\,,
 \label{DualStates}
\end{equation}
where $N_{++} + k  N_{00}  =N \equiv N_1 N_5$ and the subscript, $p$, on $|\dots\rangle_p$ indicates the strand length.%
\footnote{As usual, one must add the caveat that the language for describing these states is adapted to the symmetric orbifold locus in the moduli space of the D1-D5 system, which robustly characterizes the two-charge BPS ground states in the Ramond sector of this dual CFT.  The utility of this language for describing generic excited states above these ground states is unclear, since the regime of the moduli space where supergravity is the effective low-energy approximation is far from the symmetric orbifold regime, but this characterization has proved quite useful for characterizing the particular three-charge microstates constructed in~\cite{Bena:2015bea,Bena:2016ypk,Bena:2017xbt}.}

The microstate geometries dual to such coherent states were first presented in \cite{Bena:2016ypk}.   Apart from the quantum numbers $N_1, N_5$ and $k$, there are two Fourier coefficients, $a$ and $b$, that determine the numbers of $|\!+\!+\rangle_1$ and $|00\rangle_k$ strands, respectively.  In the supergravity theory, the partitioning of the strands emerges as a regularity condition at the D1-D5 locus and takes the form:
\begin{equation} 
\frac{Q_1Q_5}{R_y^2} ~=~ a^2 + \coeff{1}{2}\, b^2 \,, 
\label{strandbudget1}
\end{equation} 
where $Q_I$ are the supergravity charges and $R_y$ is the radius of the common $y$-circle of the D1 and D5 branes.  Again, since we are working with the D1-D5 system, we have relabelled $Q_2 = Q_5$.  The constraint (\ref{strandbudget1}) is simply the generalization of (\ref{STreg}) to systems consisting of a mixture of  $|\!+\!+\rangle$ strands and  $| 0 0\rangle$ strands.

The supergravity charges are related to the quantized charges as in (\ref{Q1Q5_N1N5a})  and the supergravity momentum charge, $Q_P$, is related to the quantized momentum charge (along the $y$-direction) via \cite{Bena:2015bea}:
\begin{equation}
N_P ~=~    \cN \, Q_P \,.
\label{QP_NP}
\end{equation}
where $\cN$ is given by (\ref{cNdefn}).

\subsection{The family of superstrata}
\label{ss:superstrata}

In contrast to  \cite{Tyukov:2017uig}, there does not seem to be a particularly simple way of writing our family of metrics as an $S^3$ fibration over a $(2+1)$-dimensional base. We therefore cast the solution in the standard six-dimensional formulation of microstate geometries. We will use the notation and conventions in earlier work \cite{Bena:2015bea,Bena:2016ypk,Bena:2017xbt}, but we will summarize the geometric details here so as to make this paper largely self-contained.

A generic superstratum depends on the phases 
 \begin{equation}
 \hat{v}_{k,m,n} ~\equiv~ \frac{\sqrt{2}}{R_y}\,(m+n) \, v ~+~ (k-m)\, \varphi_1  ~-~ m\,\varphi_2  \,.
 \label{phase}
\end{equation}
and for the states in (\ref{DualStates}) one takes a single mode with $m=0$ and $n=1$.  

It is also convenient to introduce the functions
 \begin{equation}
 \Delta_{k,m,n}~\equiv~  \left(\frac{a}{\sqrt{r^2+a^2}}\right)^k  \left(\frac{r}{\sqrt{r^2+a^2}}\right)^n \, \sin^{k-m}\theta\,  \cos^{m}\theta \,, \qquad
 \Gamma  ~\equiv~ \frac{a^2 \,\sin^2 \theta}{(r^2 +a^2)} \,, 
 \label{DeltaGammadefn}
\end{equation}
so that 
\begin{equation}
 \Delta_{2k,0,0}  ~=~  \Gamma^{k}    \,. \label{GamDel}
\end{equation}

The electrostatic potentials, $Z_I$, and self-dual magnetic $2$-forms, $\Theta_I$ are given by 
\begin{equation}
\label{Z4Th4_solngen_O(b)}
\begin{aligned}
 Z_1& = \frac{Q_1}{\Sigma}  +\frac{ b^2 \,k\, R_y^2}{2\, Q_5}\,\frac{\Delta_{2k,0,2}}{\Sigma}\, \cos \hat{v}_{2k,0,2}  \,, \qquad Z_2 = \frac{Q_5}{\Sigma} \,, \qquad Z_4= b\,\sqrt{k}\, R_y\,\frac{\Delta_{k,0,1}}{\Sigma}\, \cos \hat{v}_{k,0,1} \,,\\
   \Theta_1 & = 0 \,,  \\
  \Theta_2 &=
- \frac{  b^2 \,k\, \sqrt{2} \, R_y }{Q_5}\, \Delta_{2k,0,2}\,
\biggl[ \left(\,r\sin\theta -\, {\Sigma\over r \sin\theta}  \right)\, \sin \hat{v}_{2k,0,2}\, \Omega^{(1)} 
    -\,  \cos \hat{v}_{2k,0,2}\, \Omega^{(3)}\, \biggr]\,,  \notag  \\
 \Theta_4 &=
- b\, \sqrt{2 k} \, \Delta_{k,0,1}
\biggl[\, \left(r\sin\theta -{\Sigma\over r \sin\theta}  \right)\, \sin \hat{v}_{k,0,1}\, \Omega^{(1)} 
   -\,  \cos \hat{v}_{k,0,1}\, \Omega^{(3)}\, \biggr]\,,
\end{aligned} 
\end{equation}
where the $\Omega^{(a)}$'s are the usual basis of self-dual $2$-forms:
\begin{equation}\label{selfdualbasis}
\begin{aligned}
\Omega^{(1)} &\equiv \frac{dr\wedge d\theta}{(r^2+a^2)\cos\theta} + \frac{r\sin\theta}{\Sigma} d \varphi_1\wedge d \varphi_2\,,\ \qquad
\Omega^{(2)} \equiv  \frac{r}{r^2+a^2} dr\wedge d \varphi_2 + \tan\theta\, d\theta\wedge d \varphi_1\,,\\
 \Omega^{(3)} &\equiv \frac{dr\wedge d \varphi_1}{r} - \cot\theta\, d\theta\wedge d \varphi_2\,.
\end{aligned}
\end{equation}
Note that these excitations have been ``coiffured.''  That is, the Fourier coefficients of $(Z_1\,,\Theta_2)$ have been locked to the squares of those of $(Z_4\,,\Theta_4)$.  This constraint is required by regularity \cite{Bena:2015bea}.
 
The remaining functions in (\ref{sixmet}) are given by:
\begin{equation}
\cP ~\equiv~   Z_1\,Z_2 -  Z_4^2 ~=~   \frac{(2 a^2 + b^2) \, R_y^2 \, \Lambda^2}{2\,\Sigma^2}   \,, \qquad \cF ~=~ - \frac{b^2}{k\,(r^2 +a^2)} \, \frac{1 - \Gamma^k}{1 -\Gamma} \,, \label{PFform}
\end{equation}
and
\begin{equation}
\Lambda ~\equiv~  \sqrt{1 ~-~  \frac{ b^2\,k\, r^2}{(2 a^2 + b^2)\,(r^2 +a^2)}\, \Gamma^k }   \,. \label{LamDefn}
\end{equation}
Note that 
\begin{equation}
\Sigma ~=~   (r^2 +a^2) (1 -\Gamma)  \,, \label{SigGamreln}
\end{equation}
and hence  $\cF$ is smooth on the supertube locus, $r=0, \theta = \frac{\pi}{2}$, or $\Gamma =1$.

It is convenient to define
\begin{equation}
H_1(\Gamma) ~\equiv~ \frac{1 - \Gamma^k}{1 -\Gamma}  \,, \qquad H_2(\Gamma) ~\equiv~ \frac{d}{d \Gamma} \, H_1(\Gamma) ~=~  \frac{1 - \Gamma^k}{(1 -\Gamma)^2}  - \frac{k\,  \Gamma^{k-1}}{(1 -\Gamma)} \,.
\label{Hdefs}
\end{equation}
One should also note that these functions are smooth on the supertube locus.  

Finally, the angular momentum components are given by
\begin{equation}
\omega ~=~ \omega_0  ~+~ \frac{b^2 \, R_y }{ \sqrt{2}\,\Sigma}\, \Big[\, \Gamma^k ~+~\frac{a^2 \sin^2 \theta  \, \cos^2 \theta}{k\, (r^2 +a^2)} \,  H_2(\Gamma)\,  \Big]\, d \varphi_1~-~ \frac{ a^2 \, b^2 \,  R_y\, r^2\, \sin^2 \theta  \, \cos^2 \theta }{ \sqrt{2}\,k\,\Sigma\, (r^2 +a^2)^2}\, H_2(\Gamma)\,  d \varphi_2 \,,
\label{angmom}
\end{equation}
where $\omega_0$ is given by (\ref{angmom0}).

Note that, as a result of the coiffuring, the metric is independent of the phase (\ref{phase}) and only depends on the RMS value, $b^2$, of the Fourier modes.

The quantized angular momenta and momentum charges of this geometry are related to the Fourier coefficients via \cite{Bena:2016ypk}:
\begin{equation}
J ~\equiv~ J_L ~=~ J_R ~=~  \coeff{1}{2}\, \cN \,  a^2  \,, \qquad   Q_P ~=~   \frac{b^2}{2\,k}  ~~\Rightarrow~~ N_P   ~=~   \cN  \, \frac{b^2}{2\,k}\,.
\label{JandP}
\end{equation}
The identity $J_L =  J_R \sim   a^2$ reflects the fact that the only source of angular momentum is the original supertube, which has a circular profile in an $\IR^2$ of the spatial $\IR^4$ base geometry. In the dual CFT the angular momentum of the state dual to this supertube comes from the fermion zero modes on the $|\!+\!+\rangle$ strands.  The  excitations are created only on the $|00\rangle$ strands and so the momentum  is proportional to $ b^2/k$.  

The solution has five free parameters: $Q_1, Q_5, a, b, k$.  However we will fix the numbers of D1 and D5 branes underlying the CFT, which means fixing $Q_1$ and $Q_5$.  Similarly, we will fix the throat depth, or angular momentum, by choosing $b/a$ or $a$.  Combined with the  regularity condition (\ref{strandbudget1}), this fixes both $a$ and $b$ separately.  Finally, because of (\ref{JandP}), the choice of  $k$ is directly related to the  momentum charge, $Q_P$.  Thus the five parameters are uniquely fixed if one specifies $N_1, N_5, N_P$, $J_L = J_R$ and imposes regularity.

It is also useful to note that (\ref{strandbudget1}) can be rewritten as
\begin{equation} 
\frac{b^2}{a^2} ~=~ 2\,\bigg(\frac{Q_1Q_5}{a^2\, R_y^2}  - 1\bigg)  ~=~ 2\,\bigg(\frac{N_1 N_5}{\cN \, a^2}  - 1\bigg) ~=~  \frac{N_1 N_5}{J}  - 2 \,.
\label{strandbudget2}
\end{equation} 
We are going to be particularly interested in microstate geometries that closely approximate non-rotating extremal BTZ black holes and so they will be ``deep, scaling geometries,'' with $a^2  ~\ll~  b^2$. Such microstate geometries will be characterized by 
\begin{equation}
b^2 ~ \approx ~\frac{2\,Q_1Q_5}{R_y^2}  \,, \qquad \frac{a^2}{b^2} ~\approx~  \frac{J}{N_1 N_5}  \,.
\label{Scaling1}
\end{equation}
Note that $\frac{a^2}{b^2}$ controls the angular momentum of the CFT state compared to the overall central charge of the CFT. 

Finally, we note that $\frac{\partial}{\partial t} = \frac{1}{\sqrt{2}}(\frac{\partial}{\partial u}+ \frac{\partial}{\partial v})$ is a Killing vector of our metric and so, in solving the wave equation, one can separate modes according to
\begin{equation}
\Phi  ~=~ e^{i \omega t} \Psi(r, \theta, y, \varphi_1,\varphi_2)    \,.
\label{sepscalar} 
\end{equation}
This means that, just as in \cite{Tyukov:2017uig, Bena:2018bbd}, one can compute  energy of a long-wavelength fluctuation localizing around the cap and derive the energy gap of the CFT dual:
\begin{equation}
E_{gap}  ~=~ \frac{a^2}{b^2}  \,  \mu ~\sim~     \frac{\mu\, J}{N_1 N_5}  \, ,
\label{Egap} 
\end{equation}
where $\mu$ is some number of order $1$.

\subsection{Limits of the metric}
\label{ss:Limits}

The $a \to 0$ limit of the superstratum is simply the extremal BTZ metric with a round $S^3$ given in  (\ref{BTZmet}).  However, 
if one has $a \ne 0$, but small, and one retains $\beta$ and $\omega_0$ while taking  $r, Q_P \gg a^2$, the metric of the superstratum becomes a simple capped BTZ metric: The metric functions and angular momentum reduce to 
\begin{equation}
\cP  ~=~   \frac{(a^2 + \coeff{1}{2} \,b^2) \, R_y^2}{\Sigma^2 }   ~\to~   \frac{Q_1 \, Q_5}{r^4}  \,, \qquad \cF ~\to~ - \frac{b^2}{k\,r^2} ~=~ - \frac{2\,Q_P}{r^2}  \,,  \qquad \omega ~\to~ \omega_0   \,. \label{PFforminf}
\end{equation}
At the other extreme, $r=0$, the metric reduces to
\begin{align}
 ds_6^2  ~=~ &- \frac{a^2 }{\sqrt{Q_1 Q_5}} \, dt^2 ~+~ \sqrt{Q_1 Q_5} \,    \bigg[ \,\frac{dr^2}{a^2}  +d\theta^2 \bigg]~+~ \frac{a^2\, R_y^2 }{\sqrt{Q_1 Q_5}} \,  \sin^2 \theta \, \Big(d\varphi_1  - \frac{dt}{R_y}\Big)^2     \notag \\ 
 & ~+~ \frac{a^2\, R_y^2 }{\sqrt{Q_1 Q_5}} \,\cos^2 \theta \, \Big(d\varphi_2  - \frac{dy}{R_y}\Big)^2 ~-~ \frac{Q_P\, R_y^2 }{\sqrt{Q_1 Q_5}} \,\Big( \sin^4\theta\, H_2(\sin^2\theta) - k \, \sin^2 \theta \,H_1(\sin^2\theta)\Big)\,d\varphi_1^2\notag \\
 &~+~ \frac{Q_P\, R_y^2 }{\sqrt{Q_1 Q_5}} \,H_1(\sin^2\theta) \, \cos^2\theta\, \Big( d\varphi_2 -  \frac{dt+ dy}{R_y}  \Big)^2 \label{metorig} \,.
\end{align}
One can easily check that this metric is smooth as $\theta \to 0$ and $\theta \to \frac{\pi}{2}$.

Perhaps the most interesting limit of this superstratum is to take $k$ large.  One should note that one cannot take $k \to \infty$ because $k \le N_1 N_5$ and for $k = N_1 N_5$ one must take $a=0$ and set the quantized momentum to its minimum non-trivial value, $N_P=1$.   We will therefore consider a limit in which $k \sim (N_1 N_5)^\gamma$ with $0 < \gamma <1$.

In such limit, functions like $\Gamma^k$ are infinitesimally small everywhere except at the supertube locus, $(r,\theta)=(0, \frac{\pi}{2})$, where  $\Gamma=1$.  Thus, away from the supertube locus, one can set $\Gamma^k, \Gamma^{k\pm1} \to 0$.   In particular, this means keeping the $1$'s in the numerators of (\ref{Hdefs}). In fact, dropping all such terms is consistent with the BPS equations and one obtains precisely the blackened supertube solution defined by (\ref{BSTcharges})  and (\ref{BSTomega}) and with
\begin{equation}
Q_P ~=~ \frac{b^2}{2\,k}\,.
\label{QPkreln}
\end{equation}

Therefore, at large $k$, our superstratum looks very much like a  blackened BPS supertube located at $(r,\theta)=(0, \frac{\pi}{2})$.  Conversely, one can view our superstratum as resolving such a singular black object into a microstate geometry.  One should also note that  $(r,\theta)=(0, \frac{\pi}{2})$ is precisely  the original supertube locus and hence the locus of the original evanescent ergosurface.

Even though the blackened supertube has closed time-like curves, the superstratum is not only smooth but also free of CTC's.  From the perspective of the charges and the quartic invariant associated with black-hole horizon areas, one might be concerned that the closed time-like curves must necessarily appear in the blackened-supertube limit.  However, our solution dodges this bullet at large $k$ because $Q_P \sim k^{-1}$ and so the blackening charge gets smaller as $k$ becomes large.   From the supergravity perspective, the appearance of the CTC's is associated with the strongly singular behavior ($\Sigma^{-3}$) of $\omega$  in (\ref{BSTomega}).  In the superstratum, this is smoothed out via the appearances of the function $H_2(\Gamma)$ in (\ref{angmom}), and $\omega$  is finite as  $(r,\theta) \to (0, \frac{\pi}{2})$.

\section{Geodesics and probes} 
\label{sec:Geodesics}

\subsection{Radially infalling geodesics in the superstrata}
\label{ss:radialss}

Unlike the microstate geometries studied in  \cite{Bena:2017upb,Tyukov:2017uig}, the metric we have constructed in this paper  does not have a conformal Killing tensor.  However, as in  \cite{Tyukov:2017uig}, we will focus on a simple class of {\it time-like} geodesic probes: ``equatorial geodesics''  with $\theta = \frac{\pi}{2}$ and hence $\frac{d\theta}{d \tau} =0$.  One can easily check that such a restriction is consistent with the geodesic equations because of the symmetries of the metric under  $\theta \to \pi - \theta$.   For  $\theta = \frac{\pi}{2}$  the coordinate, $\varphi_2$, degenerates and so we will also  have $\frac{d\varphi_2}{d \tau} =0$.   

We take $\theta = \frac{\pi}{2}$ as opposed to $\theta =0$ because we want to consider geodesics that pass through the supertube locus.  We also want to see the strongest effects of the ubiquitous ``bump function,'' $\Gamma$, defined in (\ref{DeltaGammadefn}).  
 
The isometries guarantee the following conserved momenta\footnote{As usual with geodesics, these quantities are ``momenta per unit rest mass,'' and so their dimensions must be adjusted accordingly.}:
\begin{equation}
L_1 ~=~ {K_{(1) \mu }} \frac{dx^\mu}{d \tau} \,,  \qquad L_2 ~=~ {K_{(2) \mu }} \frac{dx^\mu}{d \tau} \,,  \qquad    P ~=~ {K_{(3)   \mu }}  \frac{dx^\mu}{d \tau} \,, \qquad E ~=~ {K_{(4)  \mu }}  \frac{dx^\mu}{d \tau}   \,,
  \label{ConsMom}
\end{equation}
where the $K_{(I)}$  are the Killing vectors: $K_{(J)}  = \frac{\partial}{\partial \varphi_J}$, $K_{(3)}  = \frac{\partial}{\partial v }$ and $K_{(4)}  = \frac{\partial}{\partial u}$.

The standard quadratic conserved quantity coming from the metric is: 
\begin{equation}
g_{\mu \nu} \, \frac{dx^\mu}{d \tau}\,  \frac{dx^\nu}{d \tau}~\equiv~ -1  \,,
  \label{MetInt}
\end{equation}
which means that $\tau$ is the proper time measured on the geodesic. 

One can now use $\frac{d\theta}{d \tau} =\frac{d\varphi_2}{d \tau}  =0$ and (\ref{ConsMom}) to determine all the velocities with the exception of $\frac{dr}{d \tau}$, however, as usual, this can be determined, up to a sign, from  (\ref{MetInt}).   Since we want to consider infall, we want $\frac{dr}{d \tau} < 0$.

To remove all the centrifugal barriers and enable the geodesic to fall from large values of $r$ down to $r=0$, one must take:
\begin{equation}
L_1 = 0\,, \qquad L_2 = 0 \,, \qquad P ~=~ E  \,.
  \label{CentBarr1}
\end{equation}
One should note that, for $r \to \infty$, one has 
\begin{equation}
\frac{d u}{d\tau}  ~=~ \frac{d v}{d\tau}  ~=~ -\frac{E\sqrt{Q_{1}Q_{5}}}{r^{2}}  \qquad \Rightarrow \qquad \frac{d t}{d\tau}  ~=~  -\frac{E\sqrt{2\,Q_{1}Q_{5}}}{r^{2}}  \,, \quad\frac{d y}{d\tau}  ~=~ 0  \,.
  \label{velinf1}
\end{equation}
Thus the particle has no $y$-velocity at infinity and, for standard time-orientations ($\frac{d t}{d\tau}>0$), one must have 
\begin{equation}
E  ~<~   0  \,.
  \label{Eneg1}
\end{equation}

Define 
\begin{equation}
\Gamma_1 ~\equiv~\Gamma\, \big|_{\theta = \frac{\pi}{2}}  ~=~   \frac{a^2}{(r^2 +a^2)} \,, \qquad    \Lambda_1 ~\equiv~\Lambda\, \big|_{\theta = \frac{\pi}{2}}  ~=~ \sqrt{1 ~-~  \frac{ b^2\,k\, a^{2 k}\, r^2 }{(2 a^2 + b^2)\,(r^2 +a^2)^{k+1}}}  \,,
  \label{Lam1defn}
\end{equation}

Using   (\ref{CentBarr1}), one finds that (\ref{MetInt}) can be reduced to
\begin{equation}
\begin{aligned}
\Big(\frac{dr}{d \tau}\Big)^2  ~= \frac{E^2  \, (r^2+ a^2)}{k\, r^4 (2\, a^2 + b^2) \, \Lambda_1^2}\,  \Big[& \,b^2\, (2 a^2 + b^2)\, (1 - \Gamma_1^k) + 4\, k \, r^4\,  \Gamma_1 + 2\,  k \, b^2 \, r^2  \\
& ~-~ \frac{k \, b^2 \,r^2}{(r^2 +a^2)} \,(b^2 +4\, a^2 + 2\,k\, r^2) \Gamma_1^k \, \Big]~-~  \frac{(r^2+ a^2)}{\sqrt{Q_1 Q_5}\, \Lambda_1} ~ \,.
\end{aligned}
  \label{Radvel1}
\end{equation}

At $r=0$, this becomes
\begin{equation}
\Big(\frac{dr}{d \tau}\Big)^2  ~=~\frac{ E^2}{2\,  a^2}\, (4\, a^2 +b^2 (k+1))~-~  \frac{a^2}{\sqrt{Q_1 Q_5}} \,,
  \label{Radvel2}
\end{equation}
and, in particular, there  is no centrifugal barrier.  These probes thus fall all the way to $r=0$.

For $r, b \gg a$, one has:
\begin{equation}
\Big(\frac{dr}{d \tau}\Big)^2  ~=~\Big(2 + \frac{b^2}{k \, r^2}\Big)\,  E^2 ~-~  \frac{ \sqrt{2} \, r^2}{R_y\,b} \,,
  \label{Radvel4}
\end{equation}
which is precisely what one obtains for similar geodesics in the BTZ metric (\ref{BTZmet}) if one uses  (\ref{strandbudget1}) with $b \gg a$.  Note that if these  radial geodesics come to a halt at $r =r_* \gg a$ then 
\begin{equation}
E^2 ~=~ \frac{ \sqrt{2} \, k \, r_*^4}{R_y\,b\, \Big(2\, k \, r_*^2 +b^2 \Big)}  \,.
  \label{Evalue1}
\end{equation}
 For $r_* > b$, and particularly for large $k$, one has 
\begin{equation}
|E|   ~\sim~ \frac{ r_*}{ \sqrt{b\, R_y}}  \,.
  \label{Evalue2}
\end{equation}
Also note that, for $b \gg a$,  the AdS$_3$ region of (\ref{sixmet}) and  (\ref{BTZmet}) starts at around $r  \ge \frac{b}{\sqrt{k}}$. 

To summarize, the geodesics that we will study are those with $\theta= \frac{\pi}{2}$, $\frac{d\theta}{d \tau} =\frac{d\varphi_2}{d \tau} = 0$.  The conserved momenta are restricted to $L_1 =L_2 =0$, $L_3  = E < 0$  while $\frac{dr}{d \tau}$ is given by the negative square root in (\ref{Radvel1}). They will start in the  asymptotic AdS$_3$ region, that is, they will have $r_* > b $  and, by construction, they will fall all the way to $r =0$.  It is evident from (\ref{Radvel2}) that such a particle will be traveling at a very high speed in the ``Lab Frame''  that is at rest at the bottom of the cap.

\subsection{Geodesics in blackened supertubes and capped BTZ geometries}
\label{ss:radialCBTZ}

For comparison, we consider the radial geodesics for the blackened supertubes and generic capped BTZ geometries.  If we restrict to radial geodesics with $\theta = \frac{\pi}{2}$, then $\Sigma \to  r^2$ and the extra terms in  (\ref{BSTomega}) also vanish.  Thus, for this class of geodesics, the radial equations for blackened supertubes and capped BTZ geometries are actually the same.  As before, because the angular momenta of the geodesic are basically zero, the radial velocity is determined by the metric via  (\ref{MetInt}), and  (\ref{Radvel1}) simplifies to
\begin{equation}
\Big(\frac{dr}{d \tau}\Big)^2  ~= \frac{2\,E^2}{  r^2  }\,  \bigg[ \,  (r^2+ a^2) \,\Big(1+\frac{Q_P}{r^2}\Big)- \frac{a^4\, R_y^2}{Q_1 Q_5} \, \bigg]  ~-~  \frac{(r^2+ a^2)}{\sqrt{Q_1 Q_5}} ~ \,.
  \label{Radvel3}
\end{equation}

We again look for the stopping point, $r_*$, of radial geodesics when $r_*^2, \sqrt{Q_1 Q_5} \gg a^2$, and one finds:
\begin{equation}
E^2 ~=~ \frac{r_*^4}{2\, (r_*^2 + Q_P)\,\sqrt{Q_1 Q_5}}  \,.
  \label{Evalue3}
\end{equation}
and hence the  magnitude of the generic $E$ for $r_* \gg  \sqrt{Q_P}$ is:
\begin{equation}
|E|   ~\sim~  \frac{ r_*}{\sqrt{2} \,(Q_1 Q_5)^{1/4}}  \,.
  \label{Evalue4}
\end{equation}
which, because of (\ref{strandbudget1}), agrees with (\ref{Evalue2}) in the limit of interest, $b\gg a$.

\subsection{Tidal forces}
\label{ss:Tides}

For a geodesic with proper velocity, $V^\mu = \frac{dx^\mu}{d \tau}$, the equation of geodesic deviation is:
\begin{equation}
A^\mu ~\equiv~ \frac{D^2 S^\mu}{d \tau^2}  ~=~ - {R^\mu}_{\nu \rho \sigma} \, V^\nu S^\rho   V^\sigma \,,
  \label{Geodev1}
\end{equation}
where $S^\rho$ is the deviation vector.  By shifting the proper time coordinates of neighboring geodesics one can arrange $S^\rho V_\rho = 0$ over the family of geodesics.  Thus $S^\rho$ is a space-like vector in the rest-frame of the geodesic observer.  One can re-scale $S^\mu$ at any one point so that $S^\mu S_\mu =1 $ and then $A^\mu$ represents the acceleration per unit distance, or the tidal stress.  The skew-symmetry of the Riemann tensor means that $ A^\mu V_\mu  =0$ and so the tidal acceleration is similarly space-like, representing the tidal stress in the rest-frame of the infalling observer with velocity, $V^\mu$.   To find the largest stress one can maximize  the norm, $\sqrt{A^\mu A_\mu}$, of $A^\mu$  over all the choices of $S^\mu$, subject to the constraint $S^\mu S_\mu =1$.  
 
We will consider the geodesics defined in the previous section.  To analyze the stress forces we  introduce what is sometimes called the ``tidal tensor:'' 
\begin{equation}
{\cA^\mu}_\rho ~\equiv~ - {R^\mu}_{\nu \rho \sigma} \, V^\nu \, V^\sigma \,,
  \label{cAdefn}
\end{equation}
and consider its norm and some of its eigenvalues and eigenvectors. In particular, we define
\begin{equation}
|\cA| ~\equiv~  \sqrt{{\cA^\mu}_\rho\, {\cA^\rho}_\mu}  \,.
  \label{cAnorm}
\end{equation}
Note that since $V^\mu = \frac{dx^\mu}{d \tau}$ is dimensionless, $\cA$ has the same dimensions as the curvature tensor, $\ell^{-2}$.

If $V^\mu$ and $S_{(a)}^\mu$, $a=1, \dots 5$, are orthonormal vectors then it is trivial to see that 
\begin{equation}
|\cA|^2 ~=~ \sum_{a =1}^5 \, {A_{(a)}}^\mu \, {A_{(a)}}{}_\mu\,, 
  \label{cAnormsq}
\end{equation}
where ${A_{(a)}}^\mu$ is given by (\ref{Geodev1}) with $S^\mu = {S_{(a)}}^\mu$.   If there is one dominant direction of maximum stress then one can adapt the basis, $ {S_{(a)}}^\mu$ to this direction and $|\cA|$ will yield this maximum stress.  For our problem,  the maximum stress is spread over multiple directions and so $|\cA|$  will give an estimate of this stress up to a numerical factor of order $1$.  

One should also note that ${\cA^\mu}_\rho$ is not generically symmetric, and so the stress cannot always be directed along the displacement directions, ${S_{(a)}}^\mu$.  

\subsection{Tidal force in capped BTZ geometries}
\label{ss:universaltidal}

Even for the relatively simple blackened supertubes and capped BTZ geometries of Section \ref{ss:BSTs}, the tidal forces involve a number of different terms and identifying the leading components is somewhat non-trivial.  In the limit $a \to 0$, one recovers the base-line result of the  extremal BTZ black hole:
\begin{equation}
|\cA|_{\rm BTZ}  ~=~ \frac{ \sqrt{2}  }{\sqrt{Q_1 Q_5} }  ~=~ \frac{\sqrt{2}}{\sqrt{N_1 N_5}} \,\frac{\sqrt{{\rm Vol} (T^4)}}{\ell_{10}^4} \,,
  \label{cAthroatBTZ}
\end{equation}
which is independent of $r$.  This exhibits the expected property of black holes in that the tidal forces become arbitrarily small as the underlying charges grow large. 

For the capped BTZ geometries there are new tidal force terms that are proportional to (positive) powers of $a$ and depend upon $r$, $E$ and all the charges.  Different terms dominate in different regimes, but we are interested in geometries with $Q_1, Q_2 , Q_P \gg a^2$ and particles that descend deep into the throat having been released from the top of the throat.    In this context,  the dominant tidal force comes from the largest power of $E$ appearing in $|\cA|$.  Indeed, we find that the leading term is
\begin{equation}
|\cA|_{\rm throat}  ~\sim~ \frac{ 4\, \sqrt{6} \, a^2 \, Q_P \,E^2 }{r^{6}}  \,\bigg( \, 1  ~+~  \frac{32\, Q_P \, R_y^2 }{3\, Q_1 Q_5}\, \bigg)^{1/2}    \,,
  \label{cAcappedBTZ}
\end{equation}
where  $E$ is given by  (\ref{Evalue4}) and starting at the top the throat means $r_* \gtrsim \sqrt{Q_P}$.  The coefficient of this term provides a lower bound on the tidal forces.  Moreover, for the superstratum the charges are related to $b$ via  (\ref{strandbudget1})  and (\ref{QPkreln}) which means that the second term  in   (\ref{cAcappedBTZ}) is simply a $\cO(1/k)$ correction to the numerical coefficient.  Either way,  expression (\ref{cAgeneric}) is a lower bound on the tidal force that  is saturated for the superstratum.  

The first important point is that this is a universal result for the leading form of the tidal force.  The dominant tidal forces  depend  only on $Q_1, Q_5, a$ and $Q_P$  and  {\it do not depend} on the details  of the microstate structure.  

If we now take $r_* =  \sqrt{Q_P}$ and $r = a^{(1-\alpha)} Q_P^{\frac{1}{2}\alpha}$ and focus on the coefficient in   (\ref{cAcappedBTZ}), we obtain 
\begin{equation}
|\cA|_{\rm throat}  ~\sim~ \frac{ 2\, \sqrt{6} }{\sqrt{Q_1 Q_5}}  \, \bigg(\frac{Q_P}{a^2}\bigg)^{2 - 3 \alpha} 
~=~   \frac{ 2\, \sqrt{6} }{\sqrt{N_1 N_5}}\, \bigg(\frac{N_P}{2\, J} \bigg)^{2 - 3 \alpha}  \,\frac{\sqrt{{\rm Vol} (T^4)}}{\ell_{10}^4} \,.  \label{cAcappedBTZval}
\end{equation}
and this is the dominant tidal force for $0 < \alpha <\frac{2}{3}$. (For $\alpha > \frac{2}{3}$ the tidal force is dominated by the BTZ result  (\ref{cAthroatBTZ}).)

Now recall that the charges and smoothness conditions  for a single mode superstratum are given by \cite{Bena:2016ypk}:
\begin{equation}
Q_P ~=~ \frac{(m+n)}{2 k}\, b^2 \,,  \qquad Q_1 \, Q_5 ~=~ R_y^2 \, (a^2 + \coeff{1}{2} \, b^2) \,.  \label{Genss}
\end{equation}
If one releases the particle from the same point\footnote{Note we have dropped the factor of $\frac{(m+n)}{2 k}$.}, $r_* = b$  in all such geometries then one has
\begin{equation}
|E|   ~\sim~\sqrt{ \frac{b}{  R_y}}  \,.
  \label{Evalue5}
\end{equation}
and  (\ref{cAcappedBTZ}) gives 
\begin{equation}
|\cA|_{\rm throat}  ~\sim~  \frac{(m+n) \,  a^2   b^3}{ k \, R_y} \, \frac{ 1}{r^{6}}  \,,
  \label{cAcappedBTZ2}
\end{equation}
where we are ignoring numerical factors of order $1$.  Taking $b \gg a$ and setting $r = a^{(1-\alpha)}b^{\alpha}$, one obtains 
\begin{equation}
|\cA|_{\rm throat}  ~\sim~  \frac{(m+n)  }{ k \,b\, R_y} \, \bigg(\frac{b^2}{a^2}\bigg)^{(2 - 3 \alpha)} ~\sim~ \frac{(m+n)  }{ k} \frac{\sqrt{{\rm Vol} (T^4)}}{\ell_{10}^4}\, \frac{\left(N_{1}N_{5}\right)^{3/2}}{J^{2}}\, \left(\frac{J}{N_{1}N_{5}}\right)^{3\alpha}\,.
\label{cAcappedBTZ3}
\end{equation}

For $k=1, m=0, n=1$, this is precisely the result obtained in \cite{Tyukov:2017uig}.  It also shows that for small values of $\frac{(m+n)}{ k}$, the same conclusions are valid.  In particular, for the longest possible deep throats, with $J \sim 1$, the tidal stresses reach the Planck scale\footnote{We are implicitly assuming the size of the four-torus to be large but finite in Planck units.} at $\alpha < 1/2$, or 
\begin{equation}
r  ~\lesssim~   \sqrt{a b}\,.
\label{rlimit}
\end{equation}

From (\ref{cAcappedBTZ3}) one also sees that large values of $k$, like $k \sim (N_{1}N_{5})^\gamma$ for $\gamma >0$,  substantially soften the tidal forces.  This fits with the intuition that larger $k$ means longer strands and smaller energy gaps. However this is not the root cause of the softening:  (\ref{Genss}) shows that increasing $k$ simply decreases the momentum charge, $Q_P$.  One should remember (\ref{cAcappedBTZ}) and (\ref{cAcappedBTZval}), which show that, for simple capped BTZ geometries, the tidal stress  are determined by charges and not by microstructure details.   

It is a feature of the single-mode superstrata constructed in~\cite{Bena:2016ypk,Bena:2017xbt} that the ``bump function" profile of the momentum-carrying wave was located at a radial position $r\sim a\sqrt{n/k}$ when $n\gg m$.  If one scales the strand length $k$, then one must also scale $n$ so as to keep $n/k$ fixed in order to keep the momentum charge $Q_P$ fixed.  All these superstrata will thus have deviations of the geometry peaked at the same redshift for a given $Q_P$, and so might be expected to have similar tidal effects roughly independent of the details of $n$ and $k$ so long as the ratio $n/k$ is kept fixed.

\subsection{Tidal stress in the superstratum}
\label{ss:sstides}

Having looked at what we believe to be the generic tidal forces of capped BTZ geometries, we now look at the detailed results for the new superstrata given in Section \ref{ss:superstrata}.   As usual, we will  drop numerical factors of order $1$ throughout this discussion. 

At $r=0$  we find: 
\begin{equation}
|\cA|_{r=0}  ~=~  \frac{ E^2 \, b^2}{6\, a^4} \, (k+1) \,\sqrt{34 k^2 -50 k + 42} \,.\label{cstube}
\end{equation}
If the probe is released from rest at $r_* \sim b$, then $|E|$ is given by  (\ref{Evalue5}). 
For $k \gg 1$, when geodesic probes arrive at the bottom of the cap the stress has a magnitude given by:
\begin{equation}
|\cA|_{r=0}  ~\sim~ \frac{ b^3 k^2}{a^4 \, R_y}  ~=~ \frac{ b^4\, k^2}{a^4} \frac{1}{b\, R_y}~\sim~  \frac{k^2\, (N_1 N_5)^{\frac{3}{2}}}{J^2} \,\frac{\sqrt{{\rm Vol} (T^4)}}{\ell_{10}^4} \,, 
  \label{cAcap}
\end{equation}
where we have used the second equation in (\ref{Scaling1}).  Note that at the bottom of the deepest scaling throats, with $J=1$, this stress is super-Planckian.  Indeed, the only way to avoid the  stringy dissolution of the probe is if the throat is relatively shallow: $J \sim k\, (N_1 N_5)^{\beta}$ for $\beta > \frac{3}{4}$.  From (\ref{Egap}), this corresponds to an energy gap of
\begin{equation} 
E_{gap} ~\sim~\frac{k}{ (N_1 N_5)^{1-\beta} } \,, \qquad   \beta > \frac{3}{4} \,.
\label{shallowgap}
\end{equation}

Note also that the tidal stress (\ref{cstube}) {\it grows} with $k^2$.  This is not surprising if one recalls that the locus $r=0, \theta = \frac{\pi}{2}$ is the location of the original supertube and, at large $k$, it looks like a blackened supertube. 

At  $r = a^{(1-\alpha)} b^{\alpha}$ there are  a vast number of terms in the expression for the tidal stress, $|\cA|$, but the dominant contribution  for  $0 < \alpha <\frac{2}{3}$,  is given by:
\begin{equation}
|\cA|_{\rm throat}  ~\sim~2\,\sqrt{\frac{ 2 \, (3k +4) }{k^{3}}}\,  \frac{ a^2 \,b^2 \,E^2 }{r^{6}}  ~\sim~  \frac{1 }{k\,b\, R_y}\,\bigg(\frac{b^2}{a^2}\bigg)^{2-3\alpha} ~\sim~ \frac{\sqrt{{\rm Vol} (T^4)}}{\ell_{10}^4}\, \frac{\left(N_{1}N_{5}\right)^{3/2}}{k \,J^{2}}\, \left(\frac{J}{N_{1}N_{5}}\right)^{3\alpha}\,.
  \label{cAthroat1}
\end{equation}

Since we have $Q_P = \frac{b^2}{2k}$, this result is completely consistent with (\ref{cAcappedBTZ}) (up to factors $\cO(1)$) and with the analysis above.  Planck-scale effects still arise long before the probe reaches the cap.  Moreover, (\ref{cAcap}) shows that concentration  of the  strands  at the locus $r=0, \theta = \frac{\pi}{2}$ produces, for large $k$, a greatly enhanced tidal stress.

\section{Discussion}
\label{sec:Disc}

The original motivation for the work was to see if the large tidal forces experienced by probes in some microstate geometries \cite{Tyukov:2017uig} might be softened in microstate geometries that are dual to CFT sectors with a lower momentum-gap.   On our first iteration, we obtained (\ref{cAcappedBTZ2}), which suggests that larger values of $k$ might indeed soften the tidal stresses.  However, it soon became apparent that the softening was rather coming from a decrease in  the momentum charge of the solution (\ref{Genss})  and that there was a more universal result  (\ref{cAgeneric}),  (\ref{cAcappedBTZ}) and (\ref{cAcappedBTZval}) for any capped BTZ throat.  This universal tidal force term depends only upon the charges of the background, the release point of the probe, the radius of the $y$-circle and the scale, $a$, of the cap (which disappears in the BTZ limit, $a \to 0$).
For the deepest possible throats this term causes the tidal forces to reach the Planck scale about half-way down the throat (in the sense of the geometric mean):
\begin{equation}
r   ~\sim~ \sqrt{ a \, b}\,.   
  \label{halfway}
\end{equation}

As in \cite{Tyukov:2017uig}, we have only considered a single-mode superstratum and one might wonder whether appropriately mixing superstratum modes with different values of $k$ could give rise to a significant momentum charge while in the same time softening the ``impact''.  
%
Once again, the universality of (\ref{cAgeneric}) combined with the results of \cite{Bena:2016ypk} for generic combinations of modes suggests that,  for any finite momentum charge, the tidal forces in superstrata will always grow large ``half-way'' down the deepest capped BTZ throats.  

This means that, for the deepest possible throats, the scrambling of infalling matter into stringy states happens a long distance away from the cap that replaces the horizon. From a broader phenomenological perspective, this perhaps not so surprising because the infalling probes we are considering are dual to extremely high energy, non-BPS UV states of the dual field theory.  The deepest possible throats lie at the opposite extreme in that they are dual to the lowest-energy states in the most twisted sector of the dual CFT.    Such states involve an extremely coherent and delicate ``preparation'' on the supergravity side, and can be easily destabilized by extremely high-energy excitations coming from the UV at the top of the throat.  Indeed it was noted in  \cite{Tyukov:2017uig} that the energy scale associated with a point ``half-way'' down the longest throat is 
\begin{equation}
E  ~\sim~ \frac{1}{\sqrt{ N_1 N_5}}\,,
  \label{Ehalfway}
\end{equation}
which is the energy gap of typical states of the D1-D5 system without momentum and is parametrically larger than the energy gap of the typical states of the D1-D5-P system. 

In passing, we would like to emphasize that the fact that scaling microstate geometries are easily destabilized by incoming probes is an expected feature for any bulk solution dual to a coherent superposition of CFT states in the typical sector. Indeed, these states are expected to have a very small mass gap, and hence to be easily excited into other states. Hence, one expects the scrambling of infalling matter to ultimately involve complicated transitions between various microstates and excitations of all the low-mass-gap CFT modes. 
The tidal disruption of point particles we found suggests an intriguing possibility of how that scrambling might take place:  the first part of the scrambling process involves stringy excitations and modes that have relatively high energy and it is then these high energy modes that relax and dissipate into the lowest-energy, typical states of the highly twisted sectors.

Another interesting consequence of our work here is to provide further insight into the apparent instabilities of microstate geometries and the final end-states of infalling matter.  It has been argued \cite{Eperon:2016cdd,Marolf:2016nwu,Eperon:2017bwq,Bianchi:2018kzy} that matter will be trapped for arbitrarily long periods of time near evanescent ergosurfaces and that this can lead to instabilities because large amounts of matter can be stored in such regions with only very little energy cost as measured at infinity.  We have argued that these are essential features (and certainly not ``bugs'')  of microstate geometries.  Moreover, we noted that this result was already anticipated in the discussion of entropy enhancement \cite{Bena:2008nh} where it was shown that there is vastly more free energy available in the phase space of fluctuations near evanescent ergosurfaces.  This also points to a resolution of the ``instability'' of   evanescent ergosurfaces:  In the classical theory one can place arbitrary amounts of matter near such a region.  In the quantum theory, the amount of matter is limited by the quantization of phase space and, as is well known in microstate geometries, this can place very significant limits on the ability of a space-time to store large amounts of microstructure \cite{Bena:2007qc,deBoer:2008zn}.  So while evanescent ergosurfaces are the center of regions that can support vast amounts of free energy, the number of states that can be stored there is necessarily limited by quantum mechanics.

Finally, it is important to recall one of the central  motivations of the microstate geometry and fuzzball programs:  the idea that string theory has vast numbers of degrees of freedom that can be used to capture black-hole microstate structure and that this can be achieved without singularities and without true horizons.%
\footnote{Horizons have a natural role in  an effective field theory limit of the full string theory.  From this perspective, the outer horizon is simply where probes get irretrievably entangled with the entropic degrees of freedom of the microstate, so that the only way back out is via Hawking radiation.  The entropic degrees of freedom, of course, should see no horizon, as they store the information and re-radiate it coherently while preserving QM unitarity. For a proposed realization of this mechanism in little string theory see \cite{Martinec:2014gka}.}

Furthermore, while one expects most of the microstate structure to be intrinsically stringy, there will also be a huge range of coherent expressions of this microstate structure that can be described in terms of the massless limit of string theory, namely within supergravity.  In particular, this means that whenever general relativity suggests the formation of a singularity then, in string theory, we should seek a better description in terms of microstate geometries, microstate solutions or fuzzballs  \cite{Bena:2013dka}.  The ``instability'' around evanescent ergosurfaces is a case in point and in this paper we have shown precisely how a blackened supertube has an (admittedly non-generic) resolution in terms of microstate geometries. In turn these microstate geometries can serve as a starting point for the motion in the phase space of all possible microstate geometries, where stringy excitations are expected to be nearby. 

The particular blackened supertube we considered was pathological in that it has closed time-like curves but it serves to illustrate the broader point that singularities that form around supertubes and other evanescent ergosurfaces can have straightforward resolutions in terms of microstate geometries. It would be interesting to construct microstate geometries that also resolve blackened supertubes with two magnetic dipole moments (which do not necessarily have closed time-like curves) or black rings around the cap region.


\section*{Acknowledgments}
\vspace{-2mm}
This work was supported in part by the DOE grants DE-SC0011687 and DE-SC0009924, by the ANR grant Black-dS-String ANR-16-CE31-0004-01 and by the John Templeton Foundation grant 61169. RW and NPW are very grateful to the IPhT of CEA-Saclay for hospitality during this project.


\begin{adjustwidth}{-1mm}{-1mm} 
\bibliographystyle{utphys}      
\bibliography{microstates}       

\end{adjustwidth}


\end{document}